\documentstyle[12 pt]{article} \textheight 8.5in \textwidth 6.5in
\title{Non-ideal Boson system in the Gaussian approximation}
\author{Paolo R.I. Tommasini \thanks{ Supported by CNPq (Conselho
Nacional de Desenvolvimento Cient\'{\i}fico e Tecnol\'ogico,
Brazil).}\\ \it Institute for Theoretical Atomic and
Molecular Physics\\ \it Harvard-Smithsonian Center for Astrophysics \\
\it Cambridge, MA 02138. \\  A.F.R de Toledo Piza \\ \it Instituto
de F\'{\i}sica, Universidade de S\~ao Paulo,\\ \it C.P. 66318,
05389-970 S\~ao Paulo, SP, Brasil}
\begin{document}
\setlength{\baselineskip}{0.9 cm} \maketitle \centerline{\Large
Abstract} We investigate ground-state and thermal properties of a
system of non-relativistic bosons interacting through repulsive,
two-body interactions in a self-consistent gaussian mean-field
approximation which consists in writing the variationally determined
density operator as the most general gaussian functional of the
quantized field operators. Finite temperature results are obtained in
a grand canonical framework. Contact is made with the results of Lee,
Yang and Huang in terms of particular truncations of the gaussian
approximation. The full gaussian approximation supports a free phase
or a thermodynamically unstable phase when contact forces and a
standard renormalization scheme are used.  When applied to a
Hamiltonian with zero range forces interpreted as an effective theory
with a high momentum cut-off, the full gaussian approximation
generates a quasi-particle spectrum having an energy gap, in conflict
with perturbation theory results.
\vskip 1cm
PACS 03.75.Fi, 05.30.Jp, 32.80Pj, 67.90.+z
\vskip 2cm 

\pagebreak

\section{Introduction}

The non-relativistic interacting Bose gas is certainly among the most
extensively investigated problems of many-body physics. Early interest
in this problem has been strongly motivated by the low temperature
properties of helium \cite{LP}. It has been sustained, on the
theoretical side, by a persisting elusiveness of the deeper nature of
the condensation process \cite{NA} and, on the experimental side, by
developments which led to therecent observation of the more artificial
Bose condensates  \cite{Rb}-\cite{Li}.

Much of the persisting uneasiness about the condensation process of a
nonideal bose gas hinges on the fact that the body of available
results relies mainly  on perturbative methods, a circumstance
which tends to be considered as a liability.  Non perturbative
approximation schemes have been developed, on the other hand, in
connection with the problem of self-interacting, relativistic Bose
fields which became relevant e.g. for inflationary models of the
universe \cite{UI}. This problem has also been used as a testing
ground for non perturbative methods which could then be applied to
more complicated systems of interacting fields. In this context, the
gaussian variational approach \cite{JI} has received considerable
attention, also in view of its relation to extended mean field methods
traditionally employed in non relativistic many-body physics
\cite{LinTh,LinP,TT,KeLin}. 

The purpose of this paper is to describe an application of the
gaussian variational approximation to the much studied problem of
interacting non relativistic bosons in order to bring about a direct
confrontation of this approach with at least part of the available
results for this problem. We use a formulation \cite{LinP} which is
very close to the standard language of non relativistic, extended mean
field approaches of the Hartree-Fock-Bogolyubov type, which can on the
other hand be directly related to the methods adopted in connection
with the Schr\"odinger representation of quantum field theory
\cite{TT}. Although for simplicity we restrict ourselves to the case
of uniform systems, extending the formulation to finite, inhomogeneous
systems such as those actually realized in the recent alkeli atom
experiments is completely straightforward \cite{LinTh} using well
known many-body techniques \cite{KTr}. The inclusion of finite
temperature effects is also straightforward in the formulation we use,
so that thermodynamic properties can be studied rather easily. We show
that several of the old results can be retrieved from {\it truncated}
versions of the full gaussian approximation, including ground state
energy and phonon spectrum. This last feature is lost when one adheres
to the full gaussian approximation, notwithstanding the fact that it
is supported theoretically by the Hugenholtz-Pines theorem
\cite{HP,GN} to all perturbative orders, a feature which has been
noted long ago by Girardeau and Arnowitz \cite{GA}. The phonon
spectrum can be recoverd in a generalized RPA treatment \cite{KP} or
by using the functional derivative method developed by Hohenberg and
Martin \cite{HM} and recently re-discussed by Griffin \cite{GG}.
Problems of thermodynamic instability develop for contact forces when
one uses a renormalization procedure in which the coupling constant is
made to approach zero from {it negative} values, which has been
considered in a relativistic context e.g. by Bardeen and Moshe
\cite{BM} and which has also been extensively discussed by Stevenson
as the ``precarious'' renormalized $\lambda\phi^4$ theory \cite{SF}
and more recently by T\"urk\"oz \cite{TT} and by Kerman and Lin
\cite{KeLin}. 

The formulation which we adopt is reviewed in Section 2, where the
formal, finite temperature equilibrium solutions for a contact
repulsive interaction are obtained. A truncated version of the
gaussian variational equations is analyzed in Section 3. In Section 4,
we deal with the full gaussian approximetion. Two different
prescriptions for dealing with the divergences are considered. The
first one involves a renormalization scheme related to that proposed
by Stevenson.  As a second, alternative scheme we treat the
Hamiltonian with contact interactions as an effective theory to be
implemented with a fixed cut-off in momentum space. Numerical results
are given in this case for properties of the different phases and
phase equilibria. Section 5 contains our conclusions.

\section{Thermal gaussian approximation}

We consider an extended, uniform and isotropic system of
non-relativistic, interacting, spinless bosons described by
the Hamiltonian (in momentum representation with periodic
boundary conditions in volume $V$)

\begin{equation}
H=\sum_{\vec{k}}^{} e(k) a_{\vec{k}}^{\dag} a_{\vec{k}}  +
\frac{\lambda}{2 V} \sum_{\vec{k_{1}} \vec{k_{2}} 
\vec{q}}^{} a_{\vec{k_{1}}+\vec{q}}^{\dag} a_{\vec{k_{2}}-
\vec{q}}^{\dag} a_{\vec{k_{2}}} a_{\vec{k_{1}}} 
\end{equation}

\noindent where \( e(k) = \frac{\hbar^{2} k^{2}}{2 m}\) is the free
particle kinetic energy, and the contact repulsive ($\lambda > 0$)
interaction between a pair of particles is $\lambda \delta( \vec{r} -
\vec{r'})$. The field operators satisfy standard boson commutation
relations. Within a grand canonical description the state of the
system is described by the density operator

\[
{\cal{F}}=\frac{1}{Z}e^{-\beta{\cal{H}}}
\]

\noindent where ${\cal{H}}=H-\mu N$, $N$ being the boson number
operator, and $Z$ the grand canonical partition function

\[
Z=Tr e^{-\beta{\cal{H}}}.
\]  

\noindent Following the variational approach of Balian and
V\'en\'eroni \cite{BV}, we look for extrema of the object

\[
f(M)=Tr e^{-M}
\]

\noindent under the constraint $M-\beta{\cal{H}}=0$, which is taken
into account through the introduction of a Lagrange multiplier matrix
$B$. This leads to the variational problem

\begin{equation}
\label{02}
\delta\Phi(M,B)=\delta Tr\left[e^{-M}+B(M-\beta {\cal{H}})\right]=0.
\end{equation}

\noindent Variation of $M$ gives $B=e^{-M}$, and elimination of $B$
from Eq. \ref{02} leads to the $M$-dependent object

\begin{equation}
\label{03}
\Phi(M,B)\rightarrow\Psi(M)=Tr\left[e^{-M}(1+M-\beta{\cal{H}})\right].
\end{equation}

Balian and V\'en\'eroni show that $Z\geq\Psi(M)$ for {\it any}
$M$. Defining $z=Tr e^{-M}$ we can write

\[
e^{-M}=z{\cal{F}}_0
\]

\noindent where now $Tr{\cal{F}}_0=1$. Substituting this in
Eq.\ref{03} and varying $z$ we find the variational expression for the
grand potential $\Omega$

\begin{equation}
\label{04}
\Omega=-\frac{1}{\beta}\ln Z\leq Tr[({\cal{H}}+KT\ln {\cal{F}}_0)
{\cal{F}}_0)]
\end{equation}

\noindent where ${\cal{F}}_0$ is an arbitrary density with unit trace,
and we used the notation $1/\beta=KT$. In Eq. \ref{04} we can in
particular identify an entropy factor as $S_0=-K\;\;Tr[{\cal{F}}_0\ln
{\cal{F}}_0]$.  

The most general gaussian approximation consists in adopting for $M$
an ansatz of the form \cite{DC}

\[
M\rightarrow\sum_{\vec{k}\vec{k}'}[A_{\vec{k}\vec{k}'}
 a^\dagger_{\vec{k}} a_{\vec{k}'} + (B_{\vec{k}\vec{k}'}
 a^\dagger_{\vec{k}} a^\dagger_{\vec{k}'} + h.c.)+(C_{\vec{k}}
 a^\dagger_{\vec{k}} +h.c.)]
\]

\noindent where the matrix $A_{\vec{k}\vec{k}'}$ is hermitean. The
quadratic form appearing in $M$ can be diagonalized by a general
canonical transformation of the Bogolyubov type, which amounts to
changing to the natural orbital representation of the extended one
boson density corresponding to the gaussian density ${\cal{F}}_0$
\cite{LinTh,LinP}.  The uniformity and isotropy assumptions we make
allow us to restrict this general ansatz so that $A_{\vec{k}\vec{k}'}$
is diagonal, $B_{\vec{k}\vec{k}'}$ vanishes unless $\vec{k}=
-\vec{k}'$ and both of these matrices and the $C_{\vec{h}}$ depend
only on the magnitudes of the momentum vectors. The diagonalization of
the quadratic form is achieved in this case by defining transformed
boson operators as

\[
\eta_{\vec{k}} = x_{k}^{\ast} b_{\vec{k}} + y_{k}^{\ast}
b_{-\vec{k}}^{\dag}
\]
\[
\eta_{\vec{k}}^{\dag} = x_{k} b_{\vec{k}}^{\dag} + y_{k}
b_{-\vec{k}}
\]

\noindent where

\[
b_{\vec{k}} = a_{\vec{k}} - \Gamma_{k}
\]

\noindent and we have used the isotropy of the uniform system to
make the c-number transformation parameters $x_{k}$, $y_{k}$ and
$\Gamma_{k}$ dependent only on the magnitude of $\vec{k}$. In order
for this transformation to be canonical we have still to impose on
the $x_{k}$ and $y_{k}$ the usual normalization condition

\begin{equation}
\label{05}
|x_{k}|^2 - |y_{k}|^2 = 1.
\end{equation} 

\noindent The trace-normalized gaussian density operator is now
written explicitly as

\begin{equation}
\label{06}
{\cal F}_{0} =  \prod_{\vec{k}} \frac{1}{1 + \nu_{k}}
\left(\frac{\nu_{k}}{1 + \nu_{k}} \right)
^{\eta_{\vec{k}}^{\dag} \eta_{\vec{k}}}.
\end{equation}

\noindent Straightforward calculation shows that

\[
Tr(\eta_{\vec{k}}^{\dag} \eta_{\vec{k'}} {\cal F}_{0})=
\nu_{k} \delta_{\vec{k} \vec{k'}}
\]

\noindent so that the $\nu_{k}$ are positive quantities corresponding
to mean occupation numbers of the $\eta$-bosons. One also finds that

\[
Tr[(x^{*}_{k}a_{\vec{k}}+y^{*}_{k}a_{-\vec{k}}^{\dag})^n {\cal F}_{0}]=
Tr[(\eta_{\vec{k}}+A_{k})^n {\cal F}_{0}]= A_{k}^n
\]

\noindent with $A_{k}=x_{k}^{*} \Gamma_{k}+y_{k}^{*} \Gamma_{k}^{*}$
so that non vanishing values of the $\Gamma_{k}$ correspond to
coherent condensates of unshifted, Bogolyubov transformed bosons. We
again invoke the system uniformity to impose

\[
\Gamma_{k}=\delta_{k,0} \Gamma_{0}
\]

\noindent in the calculations to follow.

It is important to note that the truncated density ${\cal F}_{0}$ in
general breaks the global gauge symmetry of $H$ which is responsible
for the conservation of the number of $a$-bosons. The quadratic
dispersion of the number operator $N=\sum_{\vec{k}}a_{\vec{k}}^{\dag}
a_{\vec{k}}$ in this state can in fact be obtained explicitly as

\begin{eqnarray}
\langle N^{2} \rangle - \langle N \rangle^{2} &=& 2 | \Gamma_{0} |^{2}
[| x_{0} |^{2} \nu_{0} + |y_{0}|^{2} (1 + \nu_{0})] -
2\Gamma_{0}^{\ast^{2}} x_{0} y_{0}^{\ast} (1 + 2 \nu_{0}) - \nonumber
\\ &&\nonumber \\ &&-2\Gamma_{0}^{2} y_{0} x_{0}^{\ast} (1 + 2
\nu_{0}) + |\Gamma_{0}|^{2} +\sum_{\vec{k}}[|x_{k}|^{2} \nu_{k} + (1 +
\nu_{k}) |y_{k}|^{2} ] + \nonumber \\ &&+\sum_{\vec{k}} \{
[|x_{k}|^{2} \nu_{k} + (1 + \nu_{k}) |y_{k}|^2]^{2} + |x_{k}|^{2}
|y_{k}|^{2} (1+ 2 \nu_{k})^{2} \}.  \nonumber \\ \nonumber
\end{eqnarray}

\noindent Furthermore, mean values of many-boson operators taken with
respect to ${\cal F}_{0}$ will contain no irreducible many-body parts,
so that the replacement of $\cal F$ by ${\cal F}_{0}$ amounts to a
mean field approximation. The states described by ${\cal F}_{0}$ have
therefore to be interpreted as ``intrinsic'' mean field states.

An important simplification which occurs in the case of stationary
states such as we consider here is that the transformation parameters
$x_{k}$, $y_{k}$ and $\Gamma_{0}$ can be taken to be real and we can
use a simple parametric representation that automatically satisfies
the canonicity condition. It reads

\begin{equation}
\label{07}
x_{k} = \cosh \sigma_{k}, \; \; \;  y_{k} = \sinh \sigma_{k}. 
\end{equation} 

\noindent It is then straightforward to evaluate the traces involved
in Eq.\ref{04} to obtain

\begin{eqnarray}
\label{08}
\Omega &\leq&\sum_{\vec{k}} {\left(e(k) - \mu + \frac{2 \lambda
\Gamma_{0}^{2}}{V}\right)\left[\frac{(1 + 2 \nu_{k}) \cosh 2\sigma_{k}
-1}{2} \right]}-\mu \Gamma_{0}^{2}+ \frac{\lambda \Gamma_{0}^{4}}{2 V}
\nonumber \\ &&-\frac{\lambda \Gamma_{0}^{2}}{2 V} \sum_{\vec{k}}^{}
{(1 + 2 \nu_{k})\sinh 2 \sigma_{k} }+ \frac{\lambda}{V}\left
\{\sum_{\vec{k}}\left[ \frac{(1+ 2 \nu_{k})\cosh 2\sigma_{k} -1
}{2}\right] \right\}^{2} \\ &&+\frac{\lambda}{8 V}\{ \sum_{\vec{k}}^{}
{(1 + 2 \nu_{k}) \sinh 2 \sigma_{k} }\}^{2} \nonumber \\ &&-KT
\sum_{\vec{k}}^{} [(1+\nu_{k})\ln(1+\nu_{k})-
\nu_{k}\ln\nu_{k}]. \nonumber
\end{eqnarray}

\noindent In a similar way the number fixing condition $Tr[{\cal
F}_{0}N]=\langle N \rangle$ evaluates to

\begin{equation}
\label{09}
\langle N \rangle = \Gamma_{0}^{2} + \sum_{\vec{k}}^{}
\left[\frac{(1 + 2 \nu_{k}) \cosh2 \sigma_{k}  -1}{2} \right]. 
\end{equation}

\subsection{Formal equilibrium solutions}

Equations determining the form of the truncated density
${\cal F}_{0}$ appropriate for thermal equilibrium are in
general derived by requiring that $\Omega$, eq.\ref{08}, is
stationary under arbitrary variations of $\Gamma_{0}$,
$\sigma_{k}$ and $\nu_{k}$. Variation with respect to
$\Gamma_{0}$ gives the gap equation

\begin{equation}
\Gamma_{0} \left\{\frac{2 \lambda}{V} \Gamma_{0}^{2} - 2 \mu -
\frac{\lambda}{V} \sum_{\vec{k}}\left[(1+2 \nu_{k})(\sinh 2
\sigma_{k}- 2 \cosh 2\sigma_{k}) +2\right] \right\} = 0
\end{equation}

\noindent
which, besides the trivial solution $\Gamma_{0}=0$, may also admit a
solution with a non vanishing value of $\Gamma_{0}$ obtained by
requiring that the expression in curly brackets vanishes.  This
solution involves the number constraint, Eq.\ref{09}, in addition to
the values of $\nu_{k}$ and $\sigma_{k}$, which are determined by the
remaining variational conditions on $\Omega$. In order to simplify the
algebraic work involved in the study of this class of solutions it is
convenient to use the number constraint Eq.\ref{09} to eliminate
$\Gamma_{0}$ from the right hand side of Eq.\ref{08} which then
assumes the form

\[
\Omega \leq F - \mu \langle N \rangle
\]

\noindent
with $\langle N \rangle$ given by Eq.\ref{09}. This identifies a 
free energy $F$ as 

\begin{eqnarray}
\label{11}
F &=&\sum_{\vec{k}}^{} {\left(e(k) + \lambda \rho\right)\left[\frac{(1
+ 2 \nu_{k})\cosh 2\sigma_{k} -1}{2}\right]}-\frac{\lambda \rho}{2 }
\sum_{\vec{k}}{(1 + 2\nu_{k})\sinh 2 \sigma_{k}} \nonumber\\ &
&-\frac{\lambda}{2 V}\left\{\sum_{\vec{k}}^{}\left[\frac{ (1 + 2
\nu_{k}) \cosh 2\sigma_{k} -1 }{2}\right] \right\}^{2} +
\frac{\lambda}{8 V} \{ \sum_{\vec{k}}^{} {(1 + 2 \nu_{k})\sinh 2
\sigma_{k}}\}^{2} \nonumber \\ & &+\frac{\lambda}{2 V}
\sum_{\vec{k},\vec{k'}}^{} (1 + 2 \nu_{k'})\sinh 2 \sigma_{k'}
\left[\frac{(1 + 2 \nu_{k}) \cosh 2\sigma_{k}-1}{2}\right]
+\frac{\lambda \rho^2 V}{2}\nonumber\\ & &-KT \sum_{\vec{k}}^{}
[(1+\nu_{k})\ln(1+\nu_{k})- \nu_{k}\ln\nu_{k}]
\end{eqnarray} 

\noindent Extremizing $F$ by setting derivatives with respect to
$\sigma_{k}$ and $\nu_{k} $ equal to zero one gets

\begin{eqnarray}
\label{12}
& &\tanh 2 \sigma_{k} = \\
& & \frac{\lambda \rho-\frac{\lambda}{2 V}\sum_{\vec{k}}^{}
[ (1 + 2 \nu_{k})\cosh 2\sigma_{k} -1]  - \frac{\lambda}{2 V} 
\sum_{\vec{k}}^{} {(1 +  2 \nu_{k}) \sinh 2 \sigma_{k}}}
{e(k) + \lambda \rho -\frac{\lambda}{2 V}\sum_{\vec{k}}^{}
[(1 + 2 \nu_{k}) \cosh 2\sigma_{k} -1]  + \frac{\lambda}{2 V}
\sum_{\vec{k}} {(1 + 2 \nu_{k})\sinh 2 \sigma_{k}}} \nonumber
\end{eqnarray}

\noindent and

\begin{equation}
\label{13}
\nu_{k} =\frac{1}{\{exp[\sqrt{\Delta}/{KT}]-1\}}  
\end{equation}

\noindent where

\begin{eqnarray}
\label{14}
\Delta &=& e(k)^{2} + 2 \lambda e(k)\times\nonumber\\
& &\times\{\rho -\frac{\lambda}{2 V}\sum_{\vec{k}}[(1 + 2 \nu_{k}) 
\cosh 2\sigma_{k} -1] + \frac{\lambda}{2 V} \sum_{\vec{k}} 
{(1 + 2 \nu_{k})\sinh 2 \sigma_{k}}\} \nonumber \\ 
& &+4 \lambda^{2} \{\rho - \frac{\lambda}{2 V}\sum_{\vec{k}}
[(1 + 2 \nu_{k}) \cosh 2\sigma_{k} -1] \} \times\nonumber\\
& & \times \frac{\lambda}{2 V}
\sum_{\vec{k}} {(1 + 2 \nu_{k})\sinh 2 \sigma_{k}}
\end{eqnarray}

As for the trivial solution $\Gamma_{0} = 0$, we differentiate
$\Omega$ as written in Eq.\ref{08} and get

\begin{equation}
\label{15}
\tanh 2 \sigma_{k} = \frac{-\frac{\lambda}{2 V} \sum_{\vec{k}} (1 + 2
\nu_{k}) \sinh 2 \sigma_{k}}{e(k) - \mu + \frac{\lambda}{V}
\sum_{\vec{k}}[(1 + 2 \nu_{k}) \cosh 2 \sigma_{k} - 1 ] }.
\end{equation}

Finally, we stress that the results obtained in the present section
are largely formal, as they involve divergent sums. In order to allow
for the derivation of thermodynamic properties of the different phases
they must therefore be supplemented by suitable regularization and
renormalization procedures. These will be discussed in sections 3 and
4 below.

\section{Independent $\eta$-bosons and dilute \\ system limit}

In this section we consider two different truncation schemes of the
gaussian variational equations which lead to well known results for
the low temperature properties of the interacting boson system.  The
first and most drastic truncation of the gaussian approximation
consists in neglecting all terms representing interactions between
$\eta$-bosons. The result is entirely trivial in the case
$\Gamma_{0}=0$ since this implies $\sigma_{k}=0$. All effects of the
interaction are thus discarded, giving us just the ideal gas results

\begin{equation}
\nu_{k} =\frac{1}{\{exp[\frac{e(k) -\mu}{KT}]-1\}}
\end{equation}

\noindent where $\mu$ is determined by the number constraint

\begin{equation}
\rho = \frac{1}{4 \pi^2} \int_{0}^{\infty}
\frac{k^2 dk}{ \{ \exp[\frac{e(k)-\mu}{KT} ] -1\}}.
\end{equation}

As for the solution corresponding to $\Gamma_{0} \ne 0$, discarding
interaction between $\eta$-bosons amounts to dropping all double sums
in Eq.~(9). The variational conditions on $\sigma_{k}$ and $\nu_{k}$
appear then as

\begin{equation}
\tanh 2 \sigma_{k} = \frac{\lambda \rho}{e(k) + \lambda \rho}  
\end{equation}

\noindent and

\begin{equation}
\nu_{k} = \frac{1}{e^{\frac{1}{KT} \sqrt{e(k)^{2} 
+ 2 \lambda \rho e(k)}}-1}. 
\end{equation}

\noindent When using these results for calculating $F$ we replace sums
by integrals and introduce a cut-off $\Lambda$ in the range of
integration over momenta. Ignoring terms that vanish in the limit
$\Lambda \rightarrow \infty$ we get

\begin{eqnarray}
\frac{F}{V} & = &\frac{\lambda\rho^{2}}{2}
-\frac{\lambda^{2} m \Lambda \rho^{2}}{4 \pi^{2}\hbar^{2}}
+\frac{8 m^{3/2} \lambda^{5/2} \rho^{5/2}}{15 \pi^{2}\hbar^{3}} 
\nonumber \\
& &+\frac{KT}{2 \pi^{2}} \int_{0}^{\Lambda}k^2 
\ln\left\{1-\exp\left[-\frac{\sqrt{e(k)^{2}+2e(k)\lambda\rho}}
{KT}\right]\right\}dk
\end{eqnarray}

\noindent which shows that we get a linearly divergent term
proportional to $\lambda^{2}$. This term is a leftover of the
non-normal ordered terms of $H$ which involve two $\eta$-operators.
It is therefore a direct consequence of the non-trivial nature of the
Bogolyubov transformation, and can be compensated by introducing the
Fermi pseudo-potential, where the contact interaction is replaced by

\begin{eqnarray}
V_{p}(\vec{r}-\vec{r\prime})&=&\lambda\frac{\partial}{\partial(|\vec{r}
-\vec{r\prime})|}[(|\vec{r}-\vec{r\prime}|)\delta(\vec{r}-\vec{r\prime})]
\nonumber \\ &=&\lambda\delta(\vec{r}-\vec{r\prime})+
(|\vec{r}-\vec{r\prime}|)\lambda\delta'(\vec{r}-\vec{r\prime})
\nonumber
\end{eqnarray}

\noindent which can be related, in the case of dilute, cold systems to
the two boson scattering length $a$ through  

\[
\lambda = \frac{4 \pi \hbar^{2} a}{m}.
\]

\noindent The regularized ground state energy becomes

\[
\frac{E_{0}}{V}=\frac{\lambda\rho^{2}}{2}\left(1+\frac{128}{15}
\frac{(a^{3}\rho)^{1/2}}{\pi^{1/2}}\right)
\]

\noindent and the chemical potential appears as

\begin{eqnarray}
\mu & = &\lambda\rho+\frac{4 \rho^{3/2}m^{3/2}
\lambda^{5/2}}{3 \pi^{2}\hbar^{3}}\nonumber \\
& &+\frac{1}{2 \pi^{2}} \int_{0}^{\infty} 
\frac{\lambda e(k) k^{2} dk}
{ \left\{\exp\left[ \frac{\sqrt{e(k)^{2}+2 e(k) \lambda \rho}}
{KT}\right]-1\right\} \sqrt{e(k)^{2}+2 e(k) \lambda \rho } }. 
\end{eqnarray}

These are just the results obtained by Lee, Huang and Yang \cite{KY},
by Beliaev \cite{SB} and by Hugenholtz and Pines \cite{HP} under the
assumption of a macroscopic (c-number) occupation of the zero momentum
mode. Note however that in the present formulation this is replaced by
the coherent condensate associated with $\Gamma_{0}$. In addition to
this we still have non-coherent occupation of the zero momentum mode
as given by the limit of $\nu_{k}$, Eq.~(17), for $k \rightarrow 0$.
For small, non-zero temperatures this diverges as $1 / k$ and
therefore does not contribute to the density of the system. Finally,
the values obtained for the chemical potential in the condensate
($\Gamma_{0} \ne 0$) and non-condensate ($\Gamma_{0}=0$) phases
indicate the instability of the former. This may be seen very easily
from the fact that $\mu$ is always positive in Eq.~(19) and always
negative or zero in Eq.~(15).

A possible way to circumvent this drawback in the framework of the
gaussian approximation is to perform a somewhat less drastic if more
delicate truncation of the complete variational expressions, which
relies on the fact that we are working with dilute systems. There is
little change in the case of the $\Gamma_{0}\neq 0$ phase. Eqs.~(7)
and ~(16) give the depletion

\[
\rho =  \rho_{0}+\frac{8}{3}\frac{(a^{3}\rho)^{1/2}}{\pi^{1/2}}.
\]

\noindent In the limit of a dilute system one has
$(a^{3}\rho_{0})^{1/2} \simeq(a^{3}\rho)^{1/2}\ll 1$ so that one may
replace $\rho$ by $\rho_{0}$ in Eq.~(16). With this replacement the
excitation spectrum of the truncated Hamiltonian is related to the
density of the condensate as

\[
w(k)=\sqrt{e(k)^{2}+2\lambda\rho_{0}e(k)}
\]
  
As for the phase with $\Gamma_{0}=0$, the occupation numbers $\nu_{k}$
are obtained by keeping $\sigma_{k}=0$ but with no truncation. The
chemical potential $\mu$ is determined using the constraint condition
(7) that in this case reads

\[
\rho=\frac{1}{4\pi^{2}}\int_{0}^{\infty}\frac{k^{2}dk}{\exp{\frac{e(k)
-\mu+2\lambda\rho}{KT}}-1}.
\]

\noindent Since the truncation has been avoided in the case when
$\Gamma_0=0$, this phase is no longer treated as a free bose gas, and
calculation shows that at $T=0$ the condensate phase now appears as
the stable one, since its chemical potential $\mu=\lambda\rho$ is
lower than that of the non condensed phase, for which
$\mu=2\lambda\rho$. It is apparent that in this derivation the two
phases are not treated on the same footing as far as the truncation is
concerned. However the truncation of the two body terms has a quite
different significance in each of the two cases. The rationale for
this procedure rests in fact on the expectation that the most relevant
effects of the two body interaction are incorporated to the result via
the symmetry breaking processes which take place in the treatment of
the $\Gamma_{0}\neq 0$ phase but not in that of the $\Gamma_{0}=0$
phase.

\section{Handling the full gaussian approximation }

In this section we examine two different prescriptions for dealing
with the divergences of the complete gaussian approximation developed
in section 2. The first prescription involves a renormalization scheme
similar to the one proposed by Stevenson \cite{SF} in the context of
the relativistic $\phi^{4}$ theory under the name of ``precarious
theory'', in which the bare coupling constant is made to approach zero
by negative values. It represents an attempt at sticking as much as
possible to the dynamics of contact interactions as such in the
present context. Although successful in removing the divergences in a
consistent way, this scheme will be shown to lead to a
thermodynamically unstable system.  We therefore consider also, as a
second prescription, the simple alternate ``effective theory'' scheme
in which a fixed cut-off is introduced in momentum space in the spirit
of the work of Amelino-Camelia and Pi \cite{GP}.
 
\subsection{Contact forces: precarious theory}

In order to reduce unessential complications to a minimum we restrict
ourselves in the following development to the properties of the system
at $T= 0$ in the phase corresponding to $\Gamma_{0}\ne 0$, since the
extension to $T \ne 0$ involves no additional divergences.  The gap
equation, Eq.~(8), combined with the number constraint, Eq.~(7), gives

\begin{equation}
\mu = \lambda \rho +\frac{\lambda}{2 V}\sum_{\vec{k}}
[\cosh 2\sigma_{k} -1] -\frac{\lambda}{2 V} \sum_{\vec{k}}^{}
{\sinh 2 \sigma_{k} }.  
\end{equation}

\noindent and from Eqs.~(10) to ~(12) we get

\begin{equation}
\tanh 2 \sigma_{k} = \frac{\lambda \rho -\frac{\lambda}{2 V}
\sum_{\vec{k}}[\cosh 2 \sigma_{k} -1] - \frac{\lambda}{2 V}
\sum_{\vec{k}} {\sinh 2 \sigma_{k}}}{e(k) +  \lambda \rho
-\frac{\lambda}{2 V} \sum_{\vec{k}}[\cosh 2 \sigma_{k} -1]
+ \frac{\lambda}{2 V} \sum_{\vec{k}} {\sinh 2 \sigma_{k}}}.
\end{equation}

\noindent This is an implicit equation for the transformation
parameters $\sigma_{k}$ which appear in the sums of Eqs.~(20) and
cut-off $\Lambda$ and, again neglecting contributions that vanish in
the limit $\Lambda \rightarrow \infty $, make the ansatze

\begin{equation}
\frac{1}{2 V} \sum_{\vec{k}}^{}{ \sinh 2 \sigma_{k}} =
\alpha + \beta \Lambda  
\end{equation}

\noindent and

\begin{equation}
\frac{1}{2 V} \sum_{\vec{k}}^{}{(\cosh 2\sigma_{k}-1)} = \gamma 
\end{equation}

\noindent where $\alpha$, $\beta$ and $\gamma$ are assumed to approach
finite values in the limit $\Lambda \rightarrow \infty$. We next
introduce a renormalized coupling constant $\lambda_{r}$ as

\begin{equation}
\lambda = \frac{\lambda_{r}}{1 - \frac{\lambda_{r} m \Lambda}
{2 \pi^{2} \hbar^{2}}}  
\end{equation}

\noindent so that when $\Lambda \rightarrow \infty $ the bare coupling
constant $\lambda$ approaches zero from negative values
\cite{KeLin,TT,SF}. Eq.~(21) becomes

\[
\tanh 2 \sigma_{k}=\frac{\frac{\lambda_{r} \rho-\lambda_{r}
\gamma-\lambda_{r}\alpha-\lambda_{r}\beta\Lambda} {1-\frac{\lambda_{r}
m \Lambda}{2 \pi^{2}\hbar^{2}}}} {e(k) + \frac{\lambda_{r} \rho
-\lambda_{r} \gamma + \lambda_{r} \alpha +\lambda_{r} \beta \Lambda}
{1-\frac{\lambda_{r} m \Lambda}{2 \pi^{2}\hbar^{2}}}}\equiv
-\frac{M+\frac{Q}{\Lambda}+{\cal{O}}(\Lambda^{-2})}
{e(k)+M+\frac{R}{\Lambda}+{\cal{O}}(\Lambda^{-2})}
\]

\noindent with 

\begin{eqnarray}
M&=&-\frac{2\pi^{2} \hbar^{2} \beta}{m},\nonumber \\
Q&=&\frac{2\pi^{2} \hbar^{2}}{m}(\rho-\gamma-\alpha-\frac{2\pi^{2}
\hbar^{2} \beta}{\lambda_{r} m}),\nonumber \\
R&=&-\frac{2\pi^{2} \hbar^{2}}{m}(\rho-\gamma+\alpha+\frac{2\pi^{2}
\hbar^{2} \beta}{\lambda_{r} m}).
\end{eqnarray}

\noindent This makes good sense for all $\vec{k}$ and for any
$\Lambda$ provided $M$ is positive. We can also obtain $\sinh 2
\sigma_{k}$ and $\cosh 2 \sigma_{k}$ in terms of $M$, $Q$ and $R$ to
evaluate the sums of Eqs.~(22) and ~(23) up to terms
${\cal{O}}(\Lambda^{-1})$ with the results

\begin{eqnarray}
\alpha + \beta \Lambda &=& -\frac{1}{4 \pi^{2}}
\int_{0}^{\Lambda}{\frac{k^{2}(M+\frac{Q}{\Lambda})} {\sqrt{e(k)^{2} +
2 (M+\frac{Q}{\Lambda}) e(k)+G}} dk} \nonumber \\ &=&\frac{m^{3/2}
M^{3/2}}{\pi^2 \hbar^{3}}- \frac{mQ}{2 \pi^{2} \hbar^{2}} -\frac{m M
\Lambda}{2 \pi^{2} \hbar^{2}}, \nonumber
\end{eqnarray}

\noindent which is consistent with the expression for $M$ in Eq.(25),
and

\[
\gamma = \frac{1}{4 \pi^{2}} \int_{0}^{\Lambda}
{k^2\left\{\frac{[e(k)+M+\frac{R} {\Lambda}]}{\sqrt{e(k)^{2} +
2(M+\frac{Q}{\Lambda}) e(k)+G}}-1\right\}dk} = \frac{ m^{3/2}
M^{3/2}}{3 \pi^{2} \hbar^{3}}
\]

\noindent where we defined

\[
G=\frac{2 M (R-Q)}{\Lambda}+\frac{R^{2}-Q^{2}}{\Lambda^{2}}.
\]

\noindent It is easy to see that for $M=0$ we obtain the 
ideal Bose gas results (free theory). For $M>0$, in order to evaluate 
$F$ (which for $T=0$ reduces to the ground state energy $E_{0}$), we 
also need

\[
\frac{1}{2 V} \sum_{\vec{k}}^{}{e(k)(\cosh 2\sigma_{k}-1)} = 
\frac{m M^{2} \Lambda}{4 \pi^{2} \hbar^{2}}+
\frac{mMQ}{2 \pi^{2} \hbar^{2}}-
\frac{4 m^{3/2} M^{5/2}}{5 \pi^{2} \hbar^{3}}.
\]

\noindent Taking these results to Eq.~(9) (with $\nu_{k}=0$ for $T=0$)
we see that the remaining linearly divergent terms cancel and we are
left with 

\begin{equation}
\frac{F(T=0)}{V}=\frac{E_{0}}{V}=
\frac{8 m^{3/2} M^{5/2}}{15 \pi^{2}
\hbar^{3}} - M \rho - \frac{M^{2}}{2 \lambda_{r}}
\end{equation}

\noindent which identifies the chemical potential as $\mu=-M$. Note
that $M>0$ now implies $\mu<0$. Finally, in order to relate $\mu$ (or
$M$) to the density of the system we evaluate the grand potential,
Eq.~(6), with the result

\begin{equation}
\Omega(T=0)=\frac{8m^{3/2}(-\mu)^{5/2}V}
{15\pi^{2}\hbar^{3}}-\frac{\mu^{2}V}{2\lambda_{r}}
\end{equation}

\noindent and use the relation $N=-(\frac{\partial\Omega}
{\partial\mu})_{T,V}$ to obtain

\begin{equation}
\rho=\frac{\mu}{\lambda_{r}}+\frac{4m^{3/2}
(-\mu)^{3/2}}{3\pi^{2}\hbar^{3}}.
\end{equation}

\noindent With the condition $\mu<0$ when $\Lambda \rightarrow
\infty$, we get a phonon-like spectrum as in Sec. 3,

\[
E(k) = \sqrt{e(k)^{2}-2 \mu e(k)}.
\]

\noindent We can also take the appropriate derivative of $F$ (or,
equivalently, evaluate $-\Omega / V$ using Eq.~(27)) to get the
pressure at $T=0$ as

\begin{equation}
P = -\left(\frac{\partial F}{\partial V}\right)_{T,N} =
-\frac{\Omega}{V}=
-\frac{8 m^{3/2} (-\mu^{5/2})}{15 \pi^{2} \hbar^{3}}
+\frac{\mu^{2}}{2 \lambda_{r}}
\end{equation}

\noindent and it is easy to show, using Eqs.~(26),~(28),~(29) and
$\rho>0$, that $\frac{d P}{d \rho}$ and $E_{0}/V$ are always negative.
This shows that the renormalized theory is thermodynamically unstable.
As a matter of fact, this instability should be seen as the
non-relativistic counterpart of the ``intrinsic instability'' pointed
quite some time ago by Bardeen and Moshe \cite{BM} in their analysis
of the phase-structure of the relativistic $\lambda\phi^4$ theory in
four dimensions, which has also been raised recently in the context of
multi-field theories \cite{GAC}. It underlies also the
``precariousness'' of this theory in Stevenson's treatment \cite{SF}.

Another possible phase to be studied corresponds to a solution where
$\Gamma_{0}=0$. In order to account for the fact that in this case a
macroscopic occupation $\nu_{0}$ may develop, we separate the
$\vec{k}=0$ contribution in the sums

\begin{equation}
\frac{1}{2 V} \sum_{\vec{k}}\left(\cosh 2 \sigma_{k} - 1 \right)
\rightarrow c + \frac{1}{2 V} \sum_{\vec{k}\neq 0}\left( \cosh 2
\sigma_{k} - 1 \right) \nonumber
\end{equation}

\noindent and

\begin{equation}
\frac{1}{2 V} \sum_{\vec{k}}\sinh 2 \sigma_{k} \rightarrow d +
\frac{1}{2 V} \sum_{\vec{k}\neq 0}\sinh 2 \sigma_{k} \nonumber
\end{equation}

\noindent so that Eq. (13) becomes

\begin{equation}
\tan 2 \sigma_{k} = \frac{-\lambda d - \frac{\lambda}{2 V}
\sum_{\vec{k}\neq 0} \sinh 2 \sigma_{k}}{e(k) - \mu + 2 \lambda \rho }.
\nonumber
\end{equation}

\noindent Using the ansatze (22) and (23) we obtain

\begin{equation}
\tanh 2 \sigma_{k}=\frac{\frac{-\lambda_{r}
d-\lambda_{r}\alpha-\lambda_{r}\beta\Lambda} {1-\frac{\lambda_{r} m
\Lambda}{2 \pi^{2}\hbar^{2}}}} {e(k) - \mu + \frac{2 \lambda_{r} \rho}
{1-\frac{\lambda_{r} m \Lambda}{2 \pi^{2}\hbar^{2}}}}\equiv
-\frac{M+\frac{Q}{\Lambda}+{\cal{O}}\left(\Lambda^{-2}\right)}
{e(k)-\mu+\frac{R}{\Lambda}+{\cal{O}}\left(\Lambda^{-2}\right)}
\end{equation}

\noindent with 

\begin{eqnarray}
M&=&-\frac{2\pi^{2} \hbar^{2} \beta}{m},
\nonumber \\
Q&=&+\frac{2\pi^{2} \hbar^{2}}{m} d,
\nonumber \\
R&=&-\frac{4\pi^{2} \hbar^{2}}{m} \rho. 
\end{eqnarray}

\noindent If we examine

\begin{equation}
\gamma = \frac{1}{4 \pi^{2}} \int_{0}^{\Lambda} k^{2} \left[\frac{e(k)
-\mu}{\sqrt{[e(k)-\mu]^{2}-M^{2}}} -1 \right] dk \nonumber
\end{equation}

\noindent we see that the only acceptable solution is $M=\mu$.  If
$M=0$ it follows that $\sigma_{k}=0$, leading to a possible solution
$c=d=\nu_{0}/V$ which corresponds essentially a free Bose gas. If, on
the other hand, $M\ne0$ we are again led to a thermodynamically
unstable situation.

\subsection{Effective theory}

The renormalization prescription of the preceding subsection tells us
that if $\lambda_{r}$ is positive and held fixed when $\Lambda
\rightarrow \infty$ we must have $\lambda \rightarrow 0_{-}$ which
results in a thermodynamically unstable theory that we want to
avoid. Since a positive value of $\lambda$ leads to a trivial theory
with $\lambda_{r}=0$ when the limit $\Lambda \rightarrow \infty$ is
taken, we next consider the results that one obtains for the
non-trivial effective theory in which $\Lambda$ is kept fixed at a
finite value allowing for both $\lambda_{r} > 0$ and $\lambda >
0$. This requires that one must have

\[
\frac{\lambda_{r} m \Lambda}{2 \pi^{2} \hbar^{2}} < 1
\]

\noindent and implies a finite resolution ${\cal{O}}(\Lambda^{-1})$ in
configuration space. The restricted momentum space also implies that
the validity of the results will be restricted to temperatures which
are low in the scale

\[
\frac{\hbar^{2} \Lambda^{2}}{2 K m}.
\]

\noindent Because in the derivations given in the preceding subsection
all terms that vanish in the limit $\Lambda \rightarrow \infty$ were
neglected, we use here the general expressions given in Section 2.
Eq.~(24) will be used with a fixed value of $\Lambda$ to relate the
bare and effective coupling constants $\lambda$ and $\lambda_{r}$
respectively.

Consider first the condensed phase, $\Gamma_{0}\neq 0$. The sums
appearing in Eqs.~(9),~(10) and ~(11) are now finite, and we therefore
define 

\begin{eqnarray}
A &=& \frac{1}{2 V} {\sum_{\vec{k}}^{}}' {(1 + 2 \nu_{k})\sinh 2
\sigma_{k}} \nonumber \\
B &=& \frac{1}{2 V} {\sum_{\vec{k}}^{}}' {[(1 + 2 \nu_{k})
\cosh 2\sigma_{k} -1]} \nonumber \\
C &=& \frac{1}{2 V} {\sum_{\vec{k}}^{}}' {e(k)[(1 + 2 \nu_{k})
\cosh 2\sigma_{k} -1]}. \nonumber
\end{eqnarray}

\noindent where the primed sums are restricted to $|\vec{k}|\leq
\Lambda$. Eqs.~(8), ~(10) and ~(12) appear then as

\begin{equation}
\mu = \lambda \rho -\lambda A + \lambda B
\end{equation}

\[
\tanh 2 \sigma_{\vec{k}} = \frac{\lambda [\rho -B-A]}{e(k) +
\lambda[\rho -B+A]} 
\]

\noindent and

\[
\Delta = e(k)^{2} + 2 \lambda e(k) [\rho - B + A] +
4 \lambda^{2} [\rho -B] A. 
\]

\noindent This last quantity determines $\nu_{k}$ through Eq.~(11) and
shows explicitly the presence of an energy gap in the quasiparticle
spectrum which is related to the quasiparticle interaction term $2
\lambda \rho_{0} A $ (we use $\rho-B=\rho_{0}$, the density of the
coherent condensate). The free energy $F$ and the pressure $P$ are
given by

\begin{eqnarray}
F&=&CV-\lambda \rho (A-B)V+\frac{\lambda}{2}(\rho^{2}+A^{2}-B^{2})V
+\lambda ABV \nonumber \\
& &-KT {\sum_{\vec{k}}}'[(1+\nu_{k})\ln(1+\nu_{k})-\nu_{k}
\ln\nu_{k}]
\end{eqnarray}

\noindent and

\begin{eqnarray}
P&=&\frac{\lambda B^{2}}{2} + \frac{\lambda \rho^{2}}{2} - C -\lambda
A B -\frac{\lambda A^{2}}{2} + \frac{1}{V}
{\sum_{\vec{k}}}'{\nu_{\vec{k}} \sqrt{\Delta}} \nonumber\\ & &-
\frac{KT}{V} {\sum_{\vec{k}}}'{\ln\{ 1-\exp[-\sqrt{\Delta}/KT]\}}
\end{eqnarray}

After going to the continuum limit, we can calculate $A$ and $B$ for
given values of $\rho$ and $T$ by solving numerically the system of
equations

\[
\left\{ \begin{array}{lll}
        A &=&\frac{1}{4 \pi^{2}} \int_{0}^{\Lambda}
{k^2 \{ \frac{\lambda [\rho -B-A]}{\sqrt{\Delta}}[1+\frac{2}
{\exp\frac{\sqrt{\Delta}}{KT}-1}]}\}dk   \\
        B &=&\frac{1}{4 \pi^{2}} \int_{0}^{\Lambda}
{k^2 \{ \frac{e(k)+\lambda [\rho -B+A]}{\sqrt{\Delta}}[1+\frac{2}
{\exp\frac{\sqrt{\Delta}}{KT}-1}]-1\}dk}.
        \end{array}
  \right.
\] 

\noindent Finally, in terms of the values of $A$ and $B$ we can
calculate $C$ and the thermodynamic functions also numerically.

As for the case $\Gamma_{0} = 0$ with the sums restricted to the
cut-off, it's easy to see that the only solution is $\sigma_{k}=0$. In
the continuum limit,

\begin{equation}
\rho = \frac{1}{2 \pi^{2}} \int_{0}^{\Lambda} \frac{k^{2} dk}
{\exp[(e(k)-\mu+2 \lambda \rho)/KT] -1} 
\end{equation}

\noindent which serves to determine $\mu$. The pressure is in this
case given by

\begin{equation}
P=\lambda \rho^{2}  - \frac{K T}{2 \pi^{2}} \int_{0}^{\Lambda}
{k^{2} \ln\{ 1-\exp[-(e(k)-\mu+2 \lambda \rho)/KT]\}}.
\end{equation}

\subsection{Numerical results}

\subsubsection{Cut-off dependence}

In order to perform numerical calculations we used a system of units
such that the Boltzmann constant $K=1$ and $c=1$, so that energies are
expressed in degrees Kelvin. We use $m=4\times10^{13}\;^oK$ and
$\lambda_{r} = 50 \;^oK \mbox{\AA}^{3}$. In Fig.~(1), we show the
dependence of the ground state energy, for a density $\rho=0.01
\;\mbox{\AA}^{-3}$ as the cut-off is varied in the interval $1.0
\;\mbox{\AA}^{-1} < \Lambda < 5.0\; \mbox{\AA}^{-1} $. For the
calculations described below we take $\Lambda = 3.0\; \mbox{\AA}^{-1}
$. According to Eq.~(24), this corresponds to a bare coupling constant
$\lambda=119.124^oK\mbox{\AA}^3$. For densities larger than the quoted
value one starts having stronger cut-off dependence, so that the
present scheme is, in this sense, also limited to rather dilute
systems.

\subsubsection{Phase transition}

Results of some sample calculations using the various expressions of
the preceding section are shown in Figs.~(2) to ~(5). For the range of
temperatures and densities considered here, the adopted value of the
cut-off momentum $\Lambda$ essentially saturates the values of the
finite momentum integrals involved. The density $\rho$ is plotted in
the figures as the number density in units of $\mbox{\AA}^{-3}$.

In order to study the properties of the condensed ($\rho_{0}> 0$)
and of the non-condensed ($\rho_{0}=0$) phases, as well as their
coexistence, we show in Fig.~2 the chemical potential $\mu$ calculated
as a function of the density $\rho$ in each of these two cases, for a
constant temperature fixed as $T=2^{o} K$, using Eqs.~(38) and ~(41)
respectively. The condensed phase has two different branches with
different chemical potentials in the range $\rho > \sim .7 \times
10^{-2}\mbox{\AA}^{-3}$ and ceases to exist for lower values of the
density. The chemical potential of the non-condensed phase increases
monotonically approaching the upper branch of the condensed phase
asymptotically from above. It is interesting to note that these two
solutions essentially merge at a density where the non-condensed
solution develops a macroscopic occupation $\nu_{0}$ of the zero
momentum state, and that the solution involving instead a coherent
condensate ($\rho_{0}> 0$) leads to a lower value of $\mu$. In
order to decide about the thermodynamic stability of the various
solutions we give in Fig.~3 the corresponding plot of $\mu$ against
the pressure $P$, which displays the pattern of a first order
transition. The densities of the two stable phases are indicated by
the light dashed lines in Fig.~2. Fig.~4 shows the $T=2^{o}K$ and the
$T=0^{o}K$ isotherms in a standard $P \times \rho^{-1}$ diagram
together with the appropriate Maxwell construction for $T=2^{o}K$. The
condensed solution is the only one stable at $T=0^{o}K$. The pattern
found for $T=2^{o}K$ repeats itself for higher temperatures with no
evidence of a critical temperature within the range allowed by the
limitations of the effective theory. The stable isotherms for
temperatures of 0, 2, 3 and 4 $^oK$ are shown in Fig.~5.

\section{Concluding remarks}

In the gaussian approximation it is possible to renormalize a
repulsive contact interaction using a procedure akin to the so called
precarious renormalization scheme of Stevenson\cite{SF}. Despite being
finite and exhibiting some desirable properties, such as an excitation
spectrum of the phonon type, the condensate phase is thermodynamically
unstable for all system densities. This is a  manifestation of
a corresponding instability poited out in a relativistic context by
Bardeen and Moshe \cite{BM}. Solutions with no pairing
($\sigma_{k}=0$), on the other hand, can be verified to correspond
just to a free bose gas. We stress that thethermodynamic instability
of the paired, condensed phase cannot be recognized just from the
expression for the ground state energy, Eq.~(27), but requires the use
of the equation of state.

As an alternative to such an unstable, renormalized theory we
considered a cut-off dependent effective theory for a dilute system.
Among the results of this theory we have a gap in the excitation
spectrum in disagreement with the Hugenholz-Pines theorem\cite{HP},
and a first order transition between a condensed phase and a
non-condensed phase.  The ocurence of the gap can be associated to the
drastic reordering and infinite partial ressumation of the
perturbative series implied in the Gaussian variational
approximation. It should be noted in this connection that the
variational calculation involving the truncated density $F_{0}$ is
designed to optimize the determination of the grand-potential, and
that this does not imply the variational optimization of the thermal
occupation probabilities from which the excitation spectrum is derived
\cite{BV}. On the other hand, the sensitivity of the excitation
spectrum to the dynamical ingredients included in the calculation can
be illustrated by the fact that, using the free quasi-boson truncation
described in section 3 (which breaks the variational self consistency
of the calculation), the results obtained agree with the
Hugenholtz-Pines theorem.

We also remark that it is a well known tendency of mean field
approximations to overrepresent discontinuities in situations
involving phase transitions, e.g. by predicting discontinuous
behaviors in the case of finite systems. This recommends, as already
observed in ref. \cite{GP} on the basis of a reliability analysis of
power-counting type, that caution should be exerted in the
interpretation of the first order character obtained for the phase
transition in this calculation.

In conclusion, the introduction of quantum effects at the level of a
paired mean field calculation for a non-ideal bose system gives
results which differ even qualitatively from usual perturbative
results. Comparison with experiment is still hampered by the fact
that a very dilute though sufficiently non-ideal system has not yet
been realized in practice.

\vspace{3cm}

{\large{\bf Figure Captions}}

\vspace{2cm}

\noindent FIG. 1. Cut-off dependence of the ground state energy per
particle. See text for details.

\vspace{.5cm} 

\noindent FIG. 2. Chemical potential as a funcion of number density
for T=2$^o$K and $\Lambda=3 \mbox{\AA}^{-1}$. See text for other parameter
values. The highlighted values of $\mu$ and $\rho$ correspond to the
Maxwell construction shown in Figs. 3 and 4.

\vspace{.5cm}

\noindent FIG. 3. Chemical potential as a function of pressure for the
isotherm depicted in Fig. 2. The highlighted point indicates the first
order phase transition.

\vspace{.5cm}

\noindent FIG. 4. Equation of state isotherms for T=0 and 2 $^o$K
showing the Maxwell construction identifying the phase coexistence
region in the T=2 $^o$K isotherm.

\vspace{.5cm}

\noindent FIG. 5. Equation of state stable isotherms for T=0, 2, 3 and
4 $^o$K.

\end{document}